\documentclass[10pt,letterpaper]{article}
\usepackage[top=0.85in,left=2.75in,footskip=0.75in]{geometry}

\usepackage{amsmath,amssymb}

\usepackage{changepage}

\usepackage[utf8x]{inputenc}

\usepackage{textcomp,marvosym}

\usepackage{cite}

\usepackage{nameref,hyperref}

\usepackage[right]{lineno}

\usepackage{microtype}
\DisableLigatures[f]{encoding = *, family = * }

\usepackage[table]{xcolor}

\usepackage{array}

\newcolumntype{+}{!{\vrule width 2pt}}

\usepackage[normalem]{ulem} 

\newlength\savedwidth

\newcommand\thickhline{\noalign{\global\savedwidth\arrayrulewidth\global\arrayrulewidth 2pt}%
\hline
\noalign{\global\arrayrulewidth\savedwidth}}

\raggedright
\usepackage[aboveskip=1pt,labelfont=bf,labelsep=period,justification=raggedright,singlelinecheck=off]{caption}

\usepackage{subfig}

\makeatletter
\renewcommand{\@biblabel}[1]{\quad#1.}
\makeatother

\date{}

\usepackage{lastpage,fancyhdr,graphicx}
\usepackage{epstopdf}
\fancyheadoffset[L]{2.25in}
\fancyfootoffset[L]{2.25in}

\begin{document}
\vspace*{0.2in}

\begin{flushleft}
{\Large
\textbf\newline{Evolution of urban scaling: evidence from Brazil} 
}
\newline
\\
Joao Meirelles\textsuperscript{1,2*},
Camilo Rodrigues Neto\textsuperscript{2},
Fernando Fagundes Ferreira\textsuperscript{2},
Fabiano Lemes Ribeiro\textsuperscript{3,4},
Claudia Rebeca Binder\textsuperscript{1},
\\
\bigskip
\textbf{1} Department of Civil and Environmental Engineering, Swiss Federal Institute of Technology Lausanne, Lausanne, VD, Switzerland
\\
\textbf{2} School of Arts, Sciences and Humanities, University of Sao Paulo, Sao Paulo, SP, Brazil
\\
\textbf{3} Department of Physics, Federal University of Lavras, Lavras, MG, Brazil
\\
\textbf{4} Department of Mathematics, City University of London, London, England
\\
\bigskip

%
%





* Corresponding author \\ E-mail: joao.meirelles@epfl.ch

\end{flushleft}
\section*{Abstract}
During the last years, the \textit{new science of municipalities} has been established as a fertile quantitative approach to systematically understand the urban phenomena. One of its main pillars is the proposition that urban systems display universal scaling behavior regarding socioeconomic, infrastructural and individual basic services variables. This paper discusses the extension of the universality proposition by testing it against a broad range of urban metrics in a developing country urban system. We present an exploration of the scaling exponents for over $60$ variables for the Brazilian urban system. Estimating those exponents is challenging from the technical point of view because the Brazilian municipalities' definition follows local political criteria and does not regard characteristics of the landscape, density, and basic utilities. As Brazilian municipalities can deviate significantly from urban settlements, urban-like municipalities were selected based on a systematic density cut-off procedure and the scaling exponents were estimated for this new subset of municipalities. To validate our findings we compared the results for overlaying variables with other studies based on alternative methods. It was found that the analyzed socioeconomic variables follow a \textit{superlinear} scaling relationship with the population size, and most of the infrastructure and individual basic services variables follow expected \textit{sublinear} and \textit{linear} scaling, respectively. However, some infrastructural and individual basic services variables deviated from their expected regimes, challenging the universality hypothesis of urban scaling. We propose that these deviations are a product of top-down decisions/policies. Our analysis spreads over a time-range of 10 years, what is not enough to draw conclusive observations, nevertheless we found hints that the scaling exponent of these variables are evolving towards the expected scaling regime, indicating that the deviations might be temporally constrained and that the urban systems might eventually reach the expected scaling regime.



\section*{Introduction}
%
%


Today, more than half of the world population lives in cities \cite{habitat2016urbanization} and this share is likely to increase in the next years.  Thus, it has become more and more important to understand how urban systems evolve with increasing population and what the social, economic, and ecological implications of these developments are. During the last decades, urban data had become increasingly accessible in a structured and machine-readable way. These newly available datasets, combined with approaches from complexity sciences are setting the ground for a \textit{new science of cities} \cite{west2017scale,batty2013new}. In this recent scientific approach to urban studies, a paradigm has arisen, which focuses not on the particularities of cities, but  on their common patterns. More specifically, it considers that the form and the function of urban systems are caused by universal laws that emerge from elementary local level interactions \cite{west2017scale,batty2013new,bettencourt2013origins,gomez2016explaining}. Based on this framework, recent findings have suggested similarities between cities from very different cultural, historical, geographical, and economic background, however mostly in the developed world and for a limited set of variables \cite{bettencourt2007growth, cesaretti2016population, ortman2014pre,strano2016rich}. As the possibility of formalizing a universal law of urban growth, in terms of its socio-economic and ecological impacts, might have a large implication for urban planning, it is key to understand whether this universality plays out for cities in countries with different development stages (i.e. BRICS countries, developing countries) and if the findings hold true for other variables. 

One of the foundations of this science is the scaling laws of different urban metrics \cite{west2017scale,bettencourt2013hypothesis}. In fact, during the last years, evidence has been accumulating in the scientific literature \cite{bettencourt2016urban}
 indicating that many urban variables, say $Y$, change systematically and in a nontrivial way with the population size $N$ of the city,  following the form $Y = Y_0 N^{\beta}$. Here, $Y_0$ is a constant and $\beta$ is the scaling exponent. Empirical studies suggest that in general variables related to socio-economic activities (e.g.: GDP, Patents, AIDS cases) scale in a superlinear manner with the population size ($\beta>1$) and that infrastructure variables (e.g.: pipe network, number of gas stations) scale in a sublinear manner ($\beta<1$). The findings also show that variables associated with what have been called basic individual needs (e.g.: number of houses, water consumption) scale linearly with the population of the city ($\beta=1$) \cite{bettencourt2007growth}. In this paper, we adopted the term \textit{basic individual services} to refer to such variables given the imprecision and variability of \textit{individual need} as a concept.


As these findings have been observed in different countries and years, it has been proposed that they are a universal property of cities \cite{bettencourt2013origins,gomez2016explaining,bettencourt2007growth,ortman2014pre,bettencourt2013hypothesis,bettencourt2010unified,gomez2012statistics,louf2014scaling,strano2016rich,ribeiro2017model}, which means that the proposed scaling laws would hold true for every urban system, no matter its culture, level of technology, policies, geography and so on. If this universality proposition is tested against additional data and further understood, it could bring valuable insights to urban planning processes. Although fairly consistent for diverse urban systems and across time, the universality of those scaling laws is still under dispute. Recent studies indicated scaling behavior that did not follow the proposed classification. Some studies show a high sensitivity of the scaling exponent in regards to the definition of city adopted \cite{arcaute2015constructing,louf2014smog,cottineau2017diverse,fragkias2013does}, and the statistical methods used in their estimation \cite{leitao2016scaling}. On top of that, other works indicate that the scaling exponent is sensitive to external factors, such as macroeconomic structures \cite{strano2016rich,rybski2014cities} or federal policies\cite{muller2017does}. These findings reveal a fragility in the universality hypothesis of scaling laws.

In fact it is still necessary to test the universality hypothesis of urban scaling against a greater diversity of cities around the world and for a more complete and representative series of urban indicators, as most of the published evidence comes from developed countries and from a narrow range of variables (GDP, area, street network, patents). If we are about to include this powerful framework in our future urban policies, we need to understand under which conditions it holds true. Special attention should be given to developing countries, as the lack of evidence for those urban systems is prominent and is also where most of the future urban development is expected to happen. The Brazilian Urban System is of particular interest because of its consolidated and rapid urbanization process \cite{martine2010brazil} with an increase inequality of city size over time \cite{cura2017oldnew}, which is expected to happen in other countries during the present century. The scaling behavior of Brazilian urban system had been explored in previous publications 
\cite{bettencourt2013origins,leitao2016scaling,alves2013distance,ignazzi2014scaling,ignazzi2015phd} for variables that focus mostly on socio-economic variables. Ignazzi's \cite{ignazzi2015phd} also evaluates the evolution of the scaling exponents over time with a remarkable database covering some 70 years of Brazilian urban indicators. Here, we tried to bring a different set of variables to the analysis, including mostly infrastructural and basic individual services indicators. We use previous findings as validation for our methodology. The present work intends to test the universality hypothesis of urban scaling laws by producing empirical analyses on the scaling patterns of an urban system in the developing world (Brazilian municipalities) and extending these analyzes to unexplored urban variables, mostly infrastructural variables. We also aim to investigate under which conditions urban systems deviate from the expected scaling regimes and if this deviation is ephemeral or long-lasting.


\section*{Materials and methods}

\subsection*{Variable selection}
Around $60$ variables were collected for all the $5565$ Brazilian municipalities. These variables were selected with the intention to encompass a broad range of urban domains in a descriptive way and to embrace the scaling regimes (linear, superlinear, sublinear). The variable selection was based on the hypothesis that socioeconomic productivity is superlinearly related to the population, infrastructure demand is sublinearly related to the population and basic individual services variables are linearly related to the population \cite{bettencourt2013hypothesis}. The studied variables especially focused on specific areas of the urban fabric: sanitation services, accessibility to basic services, municipality hall budget and infrastructure facilities (education, health, enterprises, etc). A complete list of variables, with the classification based on Bettencourt's proposition, is presented in the Supporting Information 

\subsection*{Data sources}
Social, economic and individual data were collected from the \textit{Brazilian Institute of Geography and Statistics} (IBGE) - coming from demographic census (IBGE-census)\cite{ibge2010censo}, from municipality government surveys (IBGE-cities)\cite{ibge2016cidades} and from the water-sewage-waste companies national survey (SNIS) \cite{snis2017serie}. The street infrastructure information was calculated from  \textit{Open Street Maps} (OSM) data \cite{osm2013brazil}. Economic indicators, revenues and expenditures variables were collected from the \textit{Brazilian Institute of Applied Economic Research} (IPEA) \cite{ipea2010atlas}. Health variables were collected from the \textit{Brazilian Institute of Unified Health Data} (DATASUS)\cite{ipea2010atlas}. Most of the data are self-declared, even for the service providers (SNIS), which brings potential biases to them. 

\subsection*{Data analysis method}
%
%

\begin{figure}[!ht]
\begin{center}
\includegraphics[width=1\textwidth]{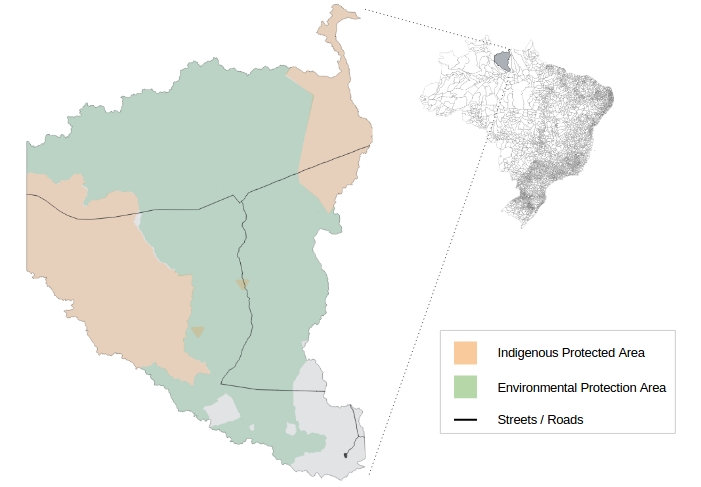}
\end{center}
\caption{{\bf Example of divergence between the definition of municipality by IBGE and the functional city definition.}
The municipality of Oriximina, located in the North of Brazil, has a geopolitical area greater than Portugal, but it has only around 50,000 inhabitants scattered around in indigenous areas and a very small urban center. The minimum density cut-offs method proposed here take off municipalities like this one from the analysis, avoiding biases that could disrupt the results. Based on data from Brazilian Ministry of Ministry of Environment\cite{mma2010} and OSM\cite{osm2013brazil}.}
\label{orixi}
\end{figure}

Urban scaling analysis should be performed from a database of cities. The spatial units of the Brazilian dataset were politically defined municipalities. This definition, however, can fundamentally differ from the functional city definition, which is a basic assumption for the non-linear scaling analysis \cite{bettencourt2013origins,ribeiro2017model}. It was previously indicated that in Brazil the municipalities themselves define their urban areas and therefore a great variability and inconsistency can be found among those definitions \cite{ignazzi2015phd}. This shows the importance for a consistent definition of urban areas in such studies for the country. A functional city is defined as a geographical area with a mixed population and high interaction intensities \cite{west2017scale}. The Brazilian definition of municipality contrast from the functional city definition as municipalities might (i) include rural characteristics with low interaction intensities between its population and (ii) be segregated from adjacent and codependent municipalities. A good example of (i) is the municipality of Oriximina, in the north of Brazil (Fig~\ref{orixi}), which has an area greater than Portugal but has a very small population of around 50,000 inhabitants and the majority of its land is composed of non-urban regions (indigenous or environmental protection areas). 

We tried so solve the problem of non-urban municipalities (i) from our dataset by adopting an approach based on the population density. Systematic subsets of the original set of municipalities were created based on minimum density cut-offs. By doing so, we excluded rural municipalities. We used the approach of excluding not dense enough locations from the original dataset, as most of the variables studied were not available at more disaggregated levels. This made it impossible to approach the problem with a generative process, which is commonly found in literature and aggregates nearby dense locations \cite{arcaute2015constructing,cottineau2017diverse}.

Regarding functional city issue (ii),  previous studies had aggregated dense, continuous and codependent municipalities into one single spatial unit. Metropolitan regions, which have been used for this purpose in other urban systems, are politically defined following different criteria in Brazil  \cite{bettencourt2007growth,bettencourt2016urban},
and a contiguity-based method would demand recognizing dense and continuous urban areas within municipalities. Since most of our indicators are provided at the city-level, any method demanding more disaggregated data becomes 
unworkable. On top of that, two different studies confronting scaling laws based on raw municipalities  \cite{alves2013distance} and metropolitan-aggregated municipalities \cite{ignazzi2014scaling} found very similar results for the same variables: superlinear scaling of GDP, unemployment, 
of education and no age change in the population. These agreements indicate that the aggregation does not really change the scaling laws found
among Brazilian municipalities. Based on those evidences, we decided not to do any aggregation of municipalities in our database and because of that it should be regarded that our findings are not related to exact functional urban areas. Even if we manage to batch the original municipalities sample to a subset of dense and urban-like municipalities, those should not be regarded as functional cities because they are not aggregated with contiguous dense surrounding areas from different municipalities.

To solve the problem of non-urban municipalities we generated systematic subsets of the municipalities based on their density and tested the scaling behavior looking for convergence. For each one of the density-thresholded subsets, ranging from 0 to 2000 $\textrm{inhabitants/}\textrm{km}^2$, the scaling exponents $\beta$, the intercept, the correlation coefficient $r^2$, and the $p-value$ of the regression were computed using ordinary least-squares (OLS) between the log-transformed population against the log-transformed of every other variable. All the regressions with p-values greater than $0.05$ or with $r^2<0.5$ were considered statistically non-significant  and therefore were ignored in the analysis. After this iterative process, one final value for the density cut-off was chosen and the results of the regression for this final dataset were analyzed. 

The list below summarizes the method adopted: 
\begin{enumerate}
\item generate subset of the original database by excluding all the municipalities with a density $\rho \leq \rho_{min}$, where $\rho_{min} \in [0, 2000]$ is a parameter of the method, which represents the minimum density of a municipality to be included in the subset;
\item fit, by ordinary least-squares (OLS), a regression between the log-transformed of the population against log-transformed of every 
other variable for all the remaining municipalities;
\item compute the scaling exponents $\beta$, the intercept, the correlation coefficient $r^2$, and the $p-value$ of the regression from the subset.
\end{enumerate}

To understand if exponents deviations changed over time, we used historical data - between 2005 and 2014 - about three specific infrastructural variables - Sewage and Water Network Length. Although data regarding previous years was available, in 2006 the sample of municipalities evaluated have almost doubled \cite{snis2012} and a very erratic patterns can be observed in the data before that. 

\section*{Results}

In this section, the results obtained from the analysis of the Brazilian municipalities dataset are presented. Given the large number of variables, only a representative sample is presented in the plots. Plots with the complete set of variables are presented in the Supporting Information \nameref{S1_Fig}, \nameref{S2_Fig} and \nameref{S2_Tab}.  

\begin{figure}[!ht]
\begin{center}
\includegraphics[width=0.8\textwidth]{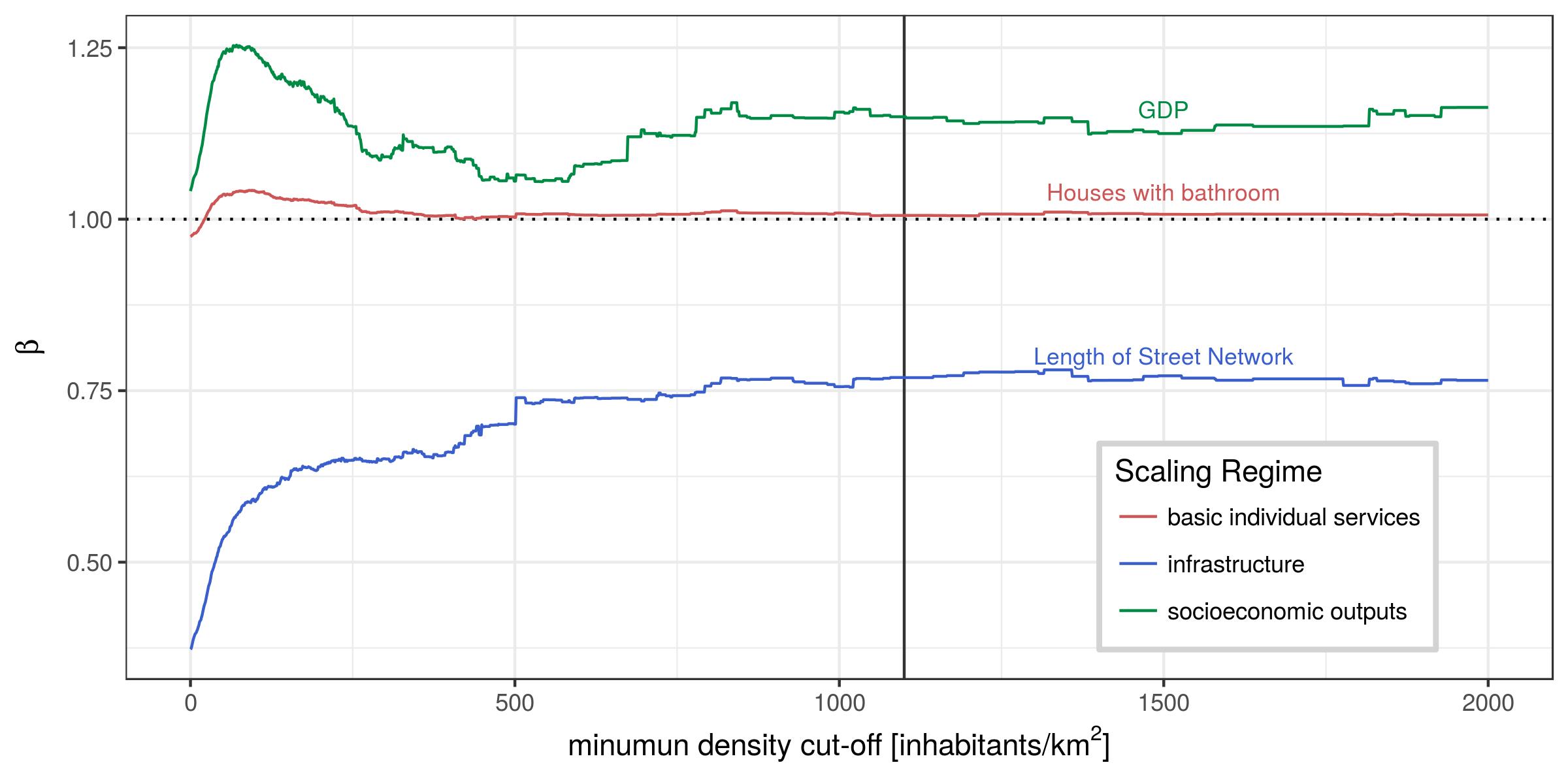}
\end{center}
\caption{{\bf Scaling exponent $\beta$ as a function of minimum density cut-off for representative variables.} Each line represents the scaling exponent (y-axis) of each variable. from OLS regressions of the log-transformed data of each variable as a function of the minimum density cut-off (x-axis). The vertical line represents a proposed final cutoff value (1100 $\textrm{inhabitants/}\textrm{km}^2$). 
}
\label{scaling_rob}
\end{figure}

In short, strong evidence of non-linear scaling in the Brazilian urban system can be found in the results. Regardless of the minimum density cutoff, most variables scale within their expected scaling regime (superlinear, sublinear, linear). This is particularly robust for the socioeconomic variables. However, some infrastructural and basic individual services variables presents a non-habitual scaling exponent across the cut-off process. This observation leads us to propose a general explanation for deviations of the scaling exponents in agreement with the urban scaling proposition and economic theory: infrastructural variables that do not emerge from local level interaction, but come from a top-down political decision and are constrained by state budgets, will tend to deviate from the expected scaling regime. As proposed by Pumain \cite{pumain2006evolutionary}, this might be part of an evolutionary process of the urban infrastructure while it does not  become a ubiquitous and universal service by serving the whole population of an urban system (here taken as a country). Though our small time-range of analysis (10 years) does not allow conclusive results. The results are detailed in the next paragraphs.
%
%
%
%
%
%
%
%
Fig~\ref{scaling_rob} shows how the scaling exponent $\beta$ changes with the minimum cut-off value of the original database for five selected variables. Each line represents one urban variable, and the colors indicate their scaling regime by Bettencourt's proposition (socio-economic, infrastructure and basic individual services variables) \cite{bettencourt2013hypothesis}. 
A version of this plot with all the statistically significant variables is presented in the Supplementary material. 

We observe a relatively high fluctuation up to a density of about $250$ $\textrm{inhabitants/}\textrm{km}^2$, what was expected given that, at this point,  the urban subset still contains a large number of non-urban municipalities. From this density cutoff value on, the scaling exponents converged. The scaling exponents show small sensitivity in relation to the \textit{city definition} in the sense that even though fluctuations are present, we observe only few changes on the coefficient regime of a variable after the very first cut-offs (i.e.: superlinear to sublinear and vice versa). This differs significantly from previous results, where the scaling exponent for other urban systems experienced a scaling regime shift \cite{louf2014scaling,arcaute2015constructing,cottineau2017diverse,fragkias2013does}. As the scaling exponent $\beta$ shows robustness in the scaling regime after the first cut-offs (around $500$ $\textrm{inhabitants/}\textrm{km}^2$), the exact point of the final cutoff will not substantially affect the exponent values.

\begin{figure}[!ht]
 \centering
\includegraphics[width=0.8\textwidth]{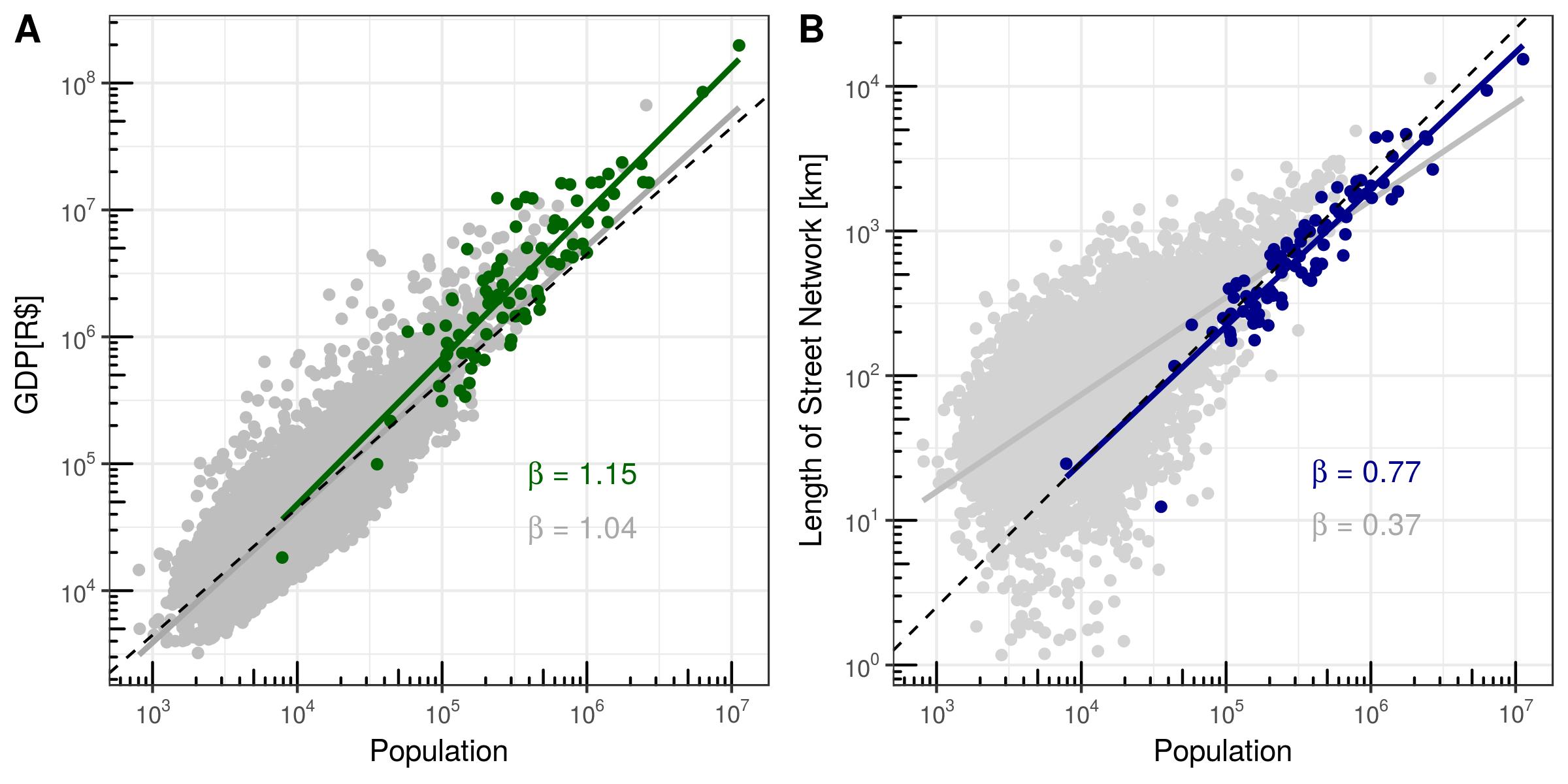}
 \caption{{\bf Scaling of representative variables for all the municipalities (gray) and for the selected ones (colored)}. Each dot represents one municipality, the x-axis indicates the population of a municipality and the y-axis indicates its variable (A - socio-economic variable: GDP; B - infrastructure variable: Length of Street Network); Colored and Gray continuous lines indicate the best-fit line from OLS regressions for the log-transformed data for each group and the black dashed line indicates the linear scaling.}
\label{scaling_final_cities}
\end{figure}

For the rest of the analyses we chose a final cutoff value at density of $1100$ $\textrm{inhabitants/}\textrm{km}^2$. This value lies within the same range of density recently adopted by OECD-EC in its city definition, which is $1500$ $\textrm{inhabitants/}\textrm{km}^2$ \cite{dijkstra2012cities}.
We opted to consider a cut-off density value smaller than the one adopted by OECD-EC just to include a greater number of municipalities. At the same time, some variables presented a small discontinuity around $1000$ $\textrm{inhabitants/}\textrm{km}^2$ and the final value was chosen to be greater than this to avoid it. Fig~\ref{scaling_final_cities} shows the final subset of 88 municipalities (colored) in comparison to the original data cloud (gray). We observe that the final set of municipalities spans over four orders of magnitude and is predominantly composed of municipalities from the upper middle deciles of population, although some small municipalities are also included. This was expected by the fact that bigger municipalities are denser and have more stabilized urban centers. 
%
%
%
%

\begin{figure}[ht]
\begin{center}
\includegraphics[width=0.8\textwidth]{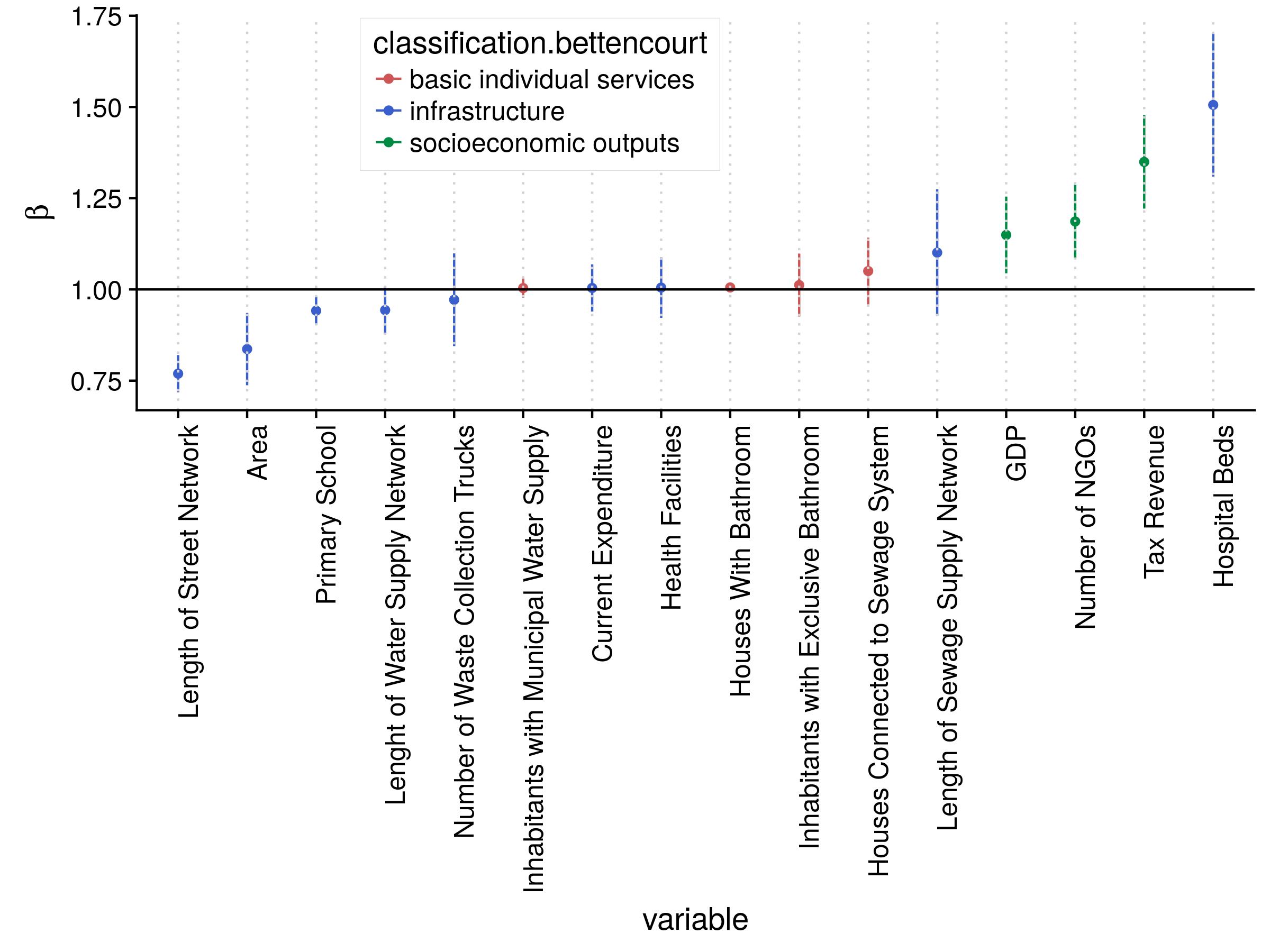}
\end{center}
\caption{{\bf Exponents values for different urban indicators in the Brazilian urban system}. Each dot represents the scaling exponent related to the best-fit line from the OLS regression of the population against the studied variable; vertical line segments represent 95\% confidence interval (CI) of those regressions; colors are based on Bettencourt's classification; the horizontal black line indicates linear relationship.}
\label{scaling_final}
\end{figure}

Fig~\ref{scaling_final} shows the final scaling exponent and confidence interval for the chosen representative set of variables. A full version of this plot, as well as a data table, containing all the statistically significant variables, can be found in the Supporting Information \nameref{S2_Tab}. An analysis of this final scaling exponents indicates that  the general proposition of increasing returns with scale holds true for our variable and methods. This can be summarized by the general scaling of revenues (superlinear) and budget (linear). However, when it comes to infrastructural or basic individual services variables, some variables  (sewage collection and treatment) deviate from their expected scaling behavior. 

\textbf{Socioeconomic variables} as GDP, as well as health-related socioeconomic outputs, as deaths by traffic accidents and homicides, scaled superlinearly. This is in accordance with previous results \cite{alves2013distance,Alves2014empiracal,ignazzi2014scaling,ignazzi2015phd}, which found superlinear scaling behavior in socioeconomic variables for the Brazilian urban system (Income, Homicides). As we had no significant deviations from earlier results, we conclude that the method here proposed is valid. Our results expand previous findings for a new set of socioeconomic variables: a diversity of different taxes revenue variables (services, urban land), social productivity (number of enterprises, length of swept sidewalks, number of NGOs). These findings are in agreement with the theoretical propositions that explain increasing return to scale of social outputs for bigger municipalities. 

The expected sublinear relationship for \textbf{infrastructure variables} was not so clear. Infrastructure variables related to general space use (area, streets), education (number of schools), water distribution (network length and number of nodes) and waste collection (number of trucks and workers) scaled sublinearly, as expected. However, variables related to sewage infrastructure or health facilities differed from their expected sublinear regimes and scaled either superlinearly (length and number of nodes of sewage network, number of hospital beds) or linearly (number of health facilities). The number of hospital beds could scale sublinearly because of the adopted definition of urban area: as we are not adopting functional areas, it could be the case that the biggest city within a urban agglomeration concentrated the hospital beds of the agglomeration and, therefore, have disproportionately more of those, leading to a superlinear scaling. If a functional area definition would be adopted, the scaling regime could change. It could also be, however, that those big municipalities concentrate hospital beds for a population beyond the urban agglomeration, receiving patients from far away municipalities as well. If the later is the case, a superlinear scaling would be expected even if a functional urban area definition would be adopted.

In relation to the \textbf{individual - basic individual services - variables}, which are expected to show a linear scaling behavior, we found some deviations as well, similar and related to the infrastructural variables. Individual services that are not related to centralized investments (number of houses with bathroom) or that are related to water supply services (access to water supply system) were found to scale linearly. This results are partly found in previous studies for the Brazilian Urban system \cite{alves2013distance}, where an aggregate index of sanitation to population access to toilets, water supply and waste collection was found to scale linearly with the population size of municipalities in Brazil. However, when it comes to sewage treatment services (number of inhabitants connected to the sewage network, the collected sewage volume, number of contracts of sewage treatment services) we found a superlinear scaling behavior. It is likely that these deviations are directly related to the deviations observed for the infrastructural variables and a possible explanation for them are presented in the Discussion section. Other variables that would be expected to follow a linear scaling behavior (i.e.: water consumption) were not statistically significant and were excluded from the analysis.


\section*{Discussion}

This research aimed at understanding whether the proposed urban scaling laws are universal, under which conditions deviations can be observed and whether these deviations are ephemeral if we test them against a broad number of variables in a economically developing urban system.

\begin{figure}[!ht]
\begin{center}
\includegraphics[width=0.9\textwidth]{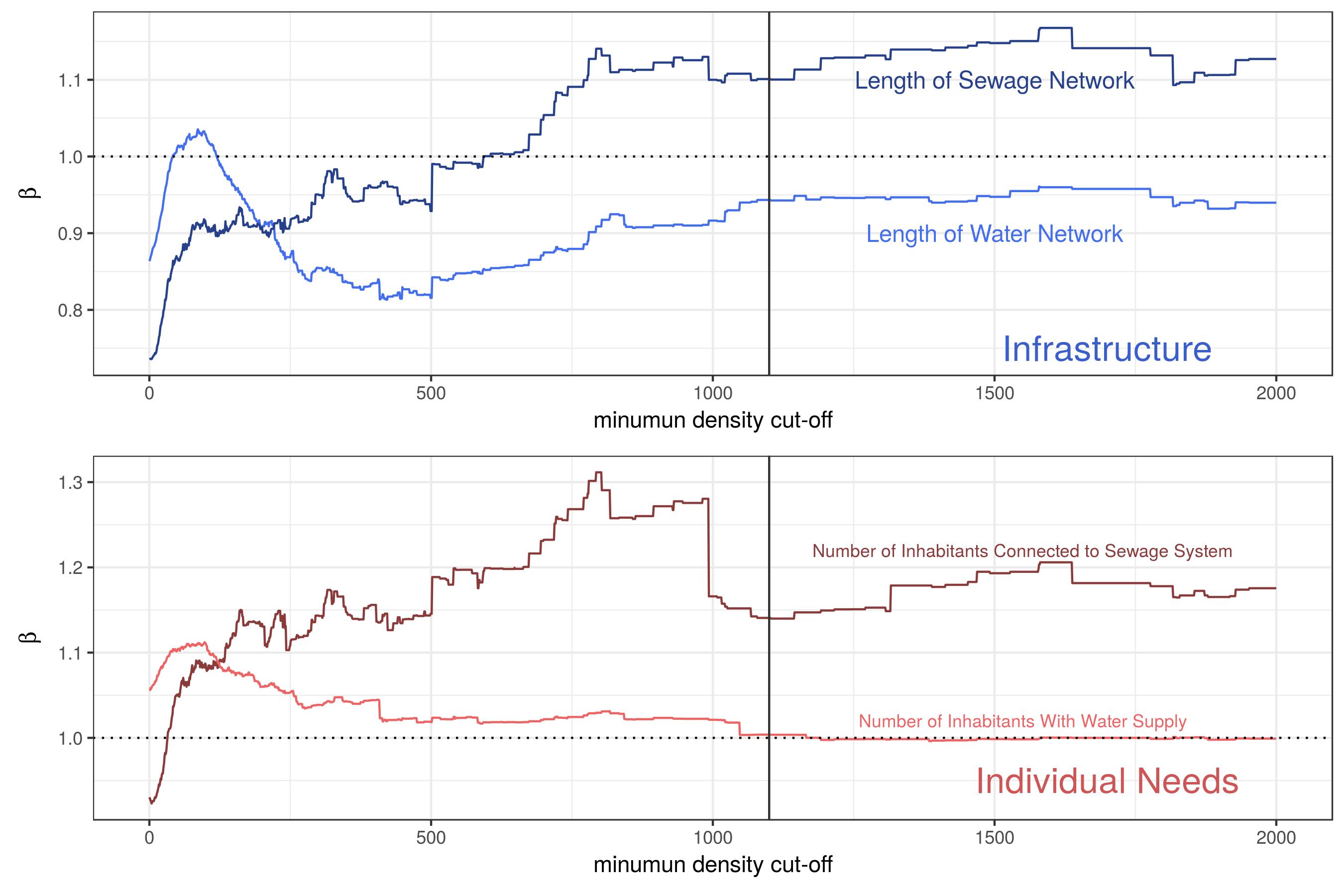}
\end{center}
\caption{{\bf Scaling exponent $\beta$ as a function of minimum density cut-off for selected variables.} Each line represents the scaling exponent (y-axis) from OLS regressions of the log-transformed data of each variable as a function of the minimum density cut-off (x-axis). A: Infrastructural Variables; B: Basic Individual Services.}
\label{scaling_rob_selected}
\end{figure}

Our results suggest that for generalizing the scaling hypothesis for infrastructural and individual variables to include developing countries some considerations have to be made. Following our results for the Brazilian urban system, not all infrastructural variables present sublinear scaling. Infrastructural variables that are provided to the whole population (water supply in the case of Brazil) and/or generated by local-level decisions (road network) do follow a sublinear scaling. The scaling regime of these cases (universal access and bottom-up generation process) is defined by social network properties and spatial constraints as proposed by Bettencourt \cite{bettencourt2013origins}. In contrast to that, infrastructural variables that depend on top-down national investments or decisions, can deviate from the sublinear regime (sewage treatment system, health facilities). This is more likely to happen in the developing world, where  the government might not be able to provide universal access to specific types of infrastructure. In this case, a centralized decision might reflect itself in a linear scaling exponent for a variable that would show a sublinear relationship if emerging from local level interactions, as is the case of sewage treatment facilities in Brazil (Fig~\ref{scaling_rob_selected}). Similarly, we can explain the shift of basic individual services variables from the expected linear exponent to a superlinear one. According to our results, in the case of Brazilian urban system, bigger municipalities with large investment possibilities are the ones which are able to guarantee infrastructure facilities to their inhabitants. 

This observation that variables without universal access (variables that are not provided to the whole population by the municipality) deviate from the expected scaling regime can be predicted from Bettencourt's \cite{bettencourt2013origins,west2017scale} theoretical framework. A necessary conditions for non-linearity to emerge is known as \textit{space filling}, which means that infrastructural variables need to have universal access in the system for the sublinear relationship of infrastructural variables to emerge. In other words, the tentacles of the distribution networks should extend everywhere in every city of the system for sublinear exponents to be observed. Given that these conditions do not hold true for sewage/health treatment infrastructure in the Brazilian urban system, it was expected that the infrastructure variables related to those services would deviate from the expected sublinear regime. The fact that some specific infrastructures are not space-filling in Brazilian municipalities suggests the necessity for further research looking into the questions whether the phenomenon observed is likely to occur in other urban systems or is a Brazil specific one. 

A economical explanation for the fact that some infrastructural variables deviate from the sublinear regime can be found in the Theory of Low-Level Equilibrium \cite{nelson1956theory}: when federal governments tend to fix prices to services that are below the economic sustainable level, decapitalizing local public facilities. Thus, cities depend on top-down, centralized interventions to expand service provision. It has been shown that sewage collection and treatment services in Brazil follow this path \cite{faria2005politicas} and, although no explicit publications were found for the economic equilibrium level of health services, it is likely that, given its high costs, the same holds true. In this scenario, only larger municipalities, with higher local tax revenues, have enough money to invest in these infrastructures. Smaller municipalities with smaller tax revenues do depend on federal investments for infrastructure expansion, as they have to spend most of their budget on staff salaries and cannot invest on the expansion of the services.  This leads to municipalities without universal access to specific types of infrastructure and, therefore, to a superlinear scaling of the variables measuring this infrastructure. The case of water distribution is different: its access was made almost universal in the country some decades ago thought public investments \cite{faria2005politicas}. The difference between variables under the low-level equilibrium regime and variables out of it can be observed in Fig~\ref{scaling_rob_selected}. Variables related to the sewage collection system deviate from their expected values, while variables related to the water supply system tend to scale within the expected range. 

\begin{figure}[ht]
\begin{center}
\includegraphics[width=0.99\textwidth]{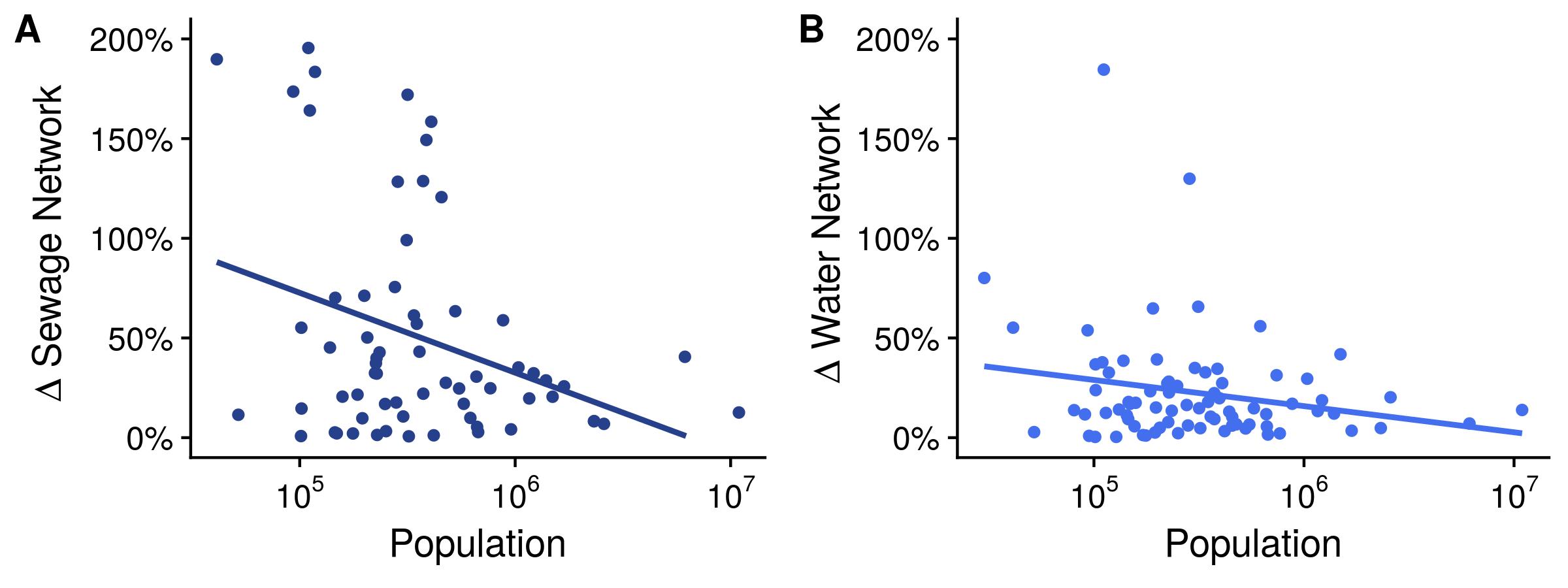}
\end{center}
\caption{{\bf Infrastructure growth between 2005 and 2014 against population in Brazilian Urban Systems.} Each dot represent the temporal variation for one municipality. A: Sewage Network; B: Water Network.}
\label{delta}
\end{figure}

One last question remains to be answered: are the observed deviations expected to continue over time or should we expect them to converge toward values found in other urban systems? Another theoretical framework can provide some insights here. Pumain \cite{pumain2006evolutionary} suggests a hierarchical diffusion process of innovations in systems of cities for variables that might become ubiquitous, making those variables disproportionately higher for larger cities until equality across cities is reached. Although this proposition has been made based on other variables, the general idea seems to hold true for the variables we studied. For our infrastructural variables, given that historically they are disproportionately higher in larger municipalities due to their low-level equilibrium, Pumains’ proposition indicates that these large municipalities would pioneer a diffusion process, adapting the technology and making it more and more affordable over time for smaller municipalities. This process had been observed before in the Brazilian urban system for different economic sectors \cite{ignazzi2015phd}. If this holds true for our infrastructural variables as well, we should observe a higher increase in the length of sewage network in smaller municipalities, and lower increase in bigger municipalities once those already have implemented the networks. To test this hypothesis, Fig~\ref{delta} presents the ten years variations on the sewage and water networks against the population of the municipalities. Although 10 years is not enough to draw conclusive observations, both infrastructural variables (water and sewage networks) seems to have increased very few in larger municipalities. Small and medium municipalities present a lot of variations, but some of them show high increases. It is also interesting to note that the sewage network have increased more than the water network, probably related to its low equilibrium situation making it a still diffusing ‘innovation’. If this dynamics is robust over greater periods of time and what drives some smaller municipalities to experience higher increases in infrastructure while other don’t remains open questions to be answered in future investigations.

\section*{Conclusion}

This paper tested the universality hypothesis of urban scaling against a broad range or variables for a urban system in a developing country. Our findings confirm that the the scaling-law holds true for the socio-economic variables. However, we found that some infrastructural and basic individual services variables (related to sewage and health services) did not scale as  proposed in literature. This observation imposes some limits to the universality hypothesis, which postulates that dense human settlements would generate the same scaling patterns regardless of the specific characteristics of the urban system. We hypothesize that the differences observed relate to country specific investment policies and economic conditions. Thus, if an infrastructural variable is dependent on top-down decisions and/or is not universally accessible across the urban systems, it can deviate from the expected sublinear scaling. In our subset of Brazilian municipalities this is the case for infrastructural variables as length of sewage collection network and health facilities, where superlinear/linear exponents were found. We also found exploratory evidences that the deviations are ephemeral and variables describing such infrastructures tend to evolve, over time, towards their expected values. These findings should be validated and new empirical evidence should be raised for other developing countries, other variables and more comprehensive functional cities definitions. Our findings show the importance of including developing countries to make the universality hypothesis truly universal.

\newpage
\section*{Acknowledgments}
We would like to acknowledge the useful and stimulating discussions with Emanuele Massaro, Emanuele Strano, Vinicius Netto and the kind review from Franziska Meinherz. 
The authors also thank the Brazilian funding agencies CAPES, CNPq and FAPESP for financial support.

\nolinenumbers

%
%
%

\newpage

\section*{Supporting information}

\paragraph*{S1 Table}
\label{S1_Tab}
{\bf Studied variables}. Description of the studied variables containing units, expected scaling regime \cite{bettencourt2013hypothesis} and source.

\begin{table}[ht!]
\centering
{\scriptsize 
\begin{tabular}{|l|c|c|c|}
\hline
{\bf Variable} & {\bf Unit}& {\bf Scaling Regime}& {\bf Source}\\ \thickhline

Population          & number        & base                & IBGE/census  \\ \hline 
Surface of Administrative Area              & Km2    & infrastructure          & IBGE/census  \\ \hline 
Gross Domestic Product                   & R\$    & social output & IPEA  \\ \hline 
Lenght of Street Network  & Km     & infrastructure          & OSM   \\ \hline 
Number of Health Facilities & number & infrastructure          & IBGE/cities  \\ \hline 
numberOfHospitalBeds      & number & infrastructure          & IBGE/cities  \\ \hline 
numberOfDaycareFacilities   & number   & infrastructure          & IBGE/cities  \\ \hline 
numberOfPrimarySchools  & number   & infrastructure          & IBGE/cities  \\ \hline 
numberOfSecondarySchools  & number & infrastructure          & IBGE/cities  \\ \hline 
numberOfNonGovernmentalOrganizations & number & social output & IBGE/cities  \\ \hline 
numberOfCommercialEnterprises & number & infrastructure & IBGE/cities  \\ \hline 
numberOfCommercialEnterprisesFacility & number & infrastructure & IBGE/cities  \\ \hline 
numberOfDeathsByTrafficAccident & number & social output & DATASUS  \\ \hline 
numberOfHomicides & number & social output & DATASUS  \\ \hline 
numberOfSuicides & number & social output & DATASUS  \\ \hline 
currentExpenditure & R\$ & infrastructure & IPEA  \\ \hline 
subsidyExpenditure & R\$ & infrastructure & IPEA  \\ \hline 
capitalExpenditure & R\$ & infrastructure & IPEA  \\ \hline 
budgetedExpenditure & R\$ & infrastructure & IPEA  \\ \hline 
expenditureByFunction & R\$ & infrastructure & IPEA  \\ \hline 
Transfer Expenditure & R\$ & social output & IPEA  \\ \hline 
budgetedRevenue & R\$ & social output & IPEA  \\ \hline 
currentRevenue & R\$ & social output & IPEA  \\ \hline 
taxRevenue & R\$ & social output & IPEA  \\ \hline 
capitalRevenue & R\$ & social output & IPEA  \\ \hline 
taxRevenueTaxes/taxes & R\$ & social output & IPEA  \\ \hline 
taxRevenueUrbanLandTax & R\$ & social output & IPEA  \\ \hline 
taxRevenueServiceTax & R\$ & social output & IPEA  \\ \hline 
taxRevenueTax- Rates & R\$ & social output & IPEA  \\ \hline 
numberOfRegisteredInhabitants & number & social output & IBGE/census  \\ \hline 
numberOfLiterateInhabitants & number & social output & IBGE/census  \\ \hline 
numberOfHousesWithBathroom & number & base & IBGE/census  \\ \hline 
numberOfHousesConnectedToSewageSystem & number & base & IBGE/census  \\ \hline 
numberOfInhabitantsWithMunicipalWaterSupply & number & base & IBGE/census  \\ \hline 
numberOfInhabitantsServedByWasteCollection & number & base & IBGE/census  \\ \hline 
numberOfInhabitantsWithAccessToElectricity & number & base & IBGE/census  \\ \hline 
numberOfInhabitantsWithExclusiveBathroom & number & base & IBGE/census  \\ \hline 
numberOfInhabitantsWithElectricityMeasurement & number & base & IBGE/census  \\ \hline 
numberOfInhabitantsWithWasteIllegalDumping & number & base & IBGE/census  \\ \hline 
numberOfInhabitantsWithWasteIllegalBurning & number & base & IBGE/census  \\ \hline 
Water and Sewage systems/Total revenue & R\$ & infrastructure & SNIS  \\ \hline 
Water and Sewage systems/Total expenditures & R\$ & infrastructure & SNIS  \\ \hline 
Water and Sewage systems/Staff expenditures & R\$ & infrastructure & SNIS  \\ \hline 
Population with water supply & number & infrastructure          & SNIS  \\ \hline 
Water suply network/links & number & infrastructure          & SNIS  \\ \hline  
Water suply network/length & Km & infrastructure          & SNIS  \\ \hline  
Water suply/electricity consumption & KhH/year & infrastructure & SNIS  \\ \hline 
Water suply/investments & R\$/year & infrastructure          & SNIS  \\ \hline 
Water suply/impacted consumers shutdowns & number & infrastructure & SNIS  \\ \hline 
Population with sewage collection & number & base & SNIS  \\ \hline 
Sewage collection network/links & number & infrastructure          & SNIS  \\ \hline 
Sewage collection network/length & Km & infrastructure          & SNIS  \\ \hline  
Sewage collection/collected volume & km3/year & base          & SNIS  \\ \hline  
Sewage collection/wastewater analyzed samples & number & infrastructure & SNIS  \\ \hline 
Waste collection/attended population & number & base          & SNIS  \\ \hline  
Waste collection/assoiated garbage collectors & number & infrastructure & SNIS  \\ \hline 
Waste collection/waste collected & Tons/year & base & SNIS  \\ \hline  
Waste collection/waste trucks & number & infrastructure & SNIS  \\ \hline  
Waste collection/total expenditures & & infrastructure & SNIS  \\ \hline  
Waste collection/total workers & number & infrastructure & SNIS  \\ \hline  
Waste collection/swept sidewalks & Km & social output & SNIS  \\ \hline  
\end{tabular}
}
\end{table}

\newpage
\paragraph*{S1 Fig}
\label{S1_Fig}

{\bf Scaling exponent $\beta$ as a function of minimum density cut-off for all the variables.} Each line represents the scaling exponent (y-axis) from OLS regressions of the log-transformed data of each variable as a function of the minimum density cut-off (x-axis).
\begin{figure}[ht!]
\begin{center}
\includegraphics[width=1\textwidth]{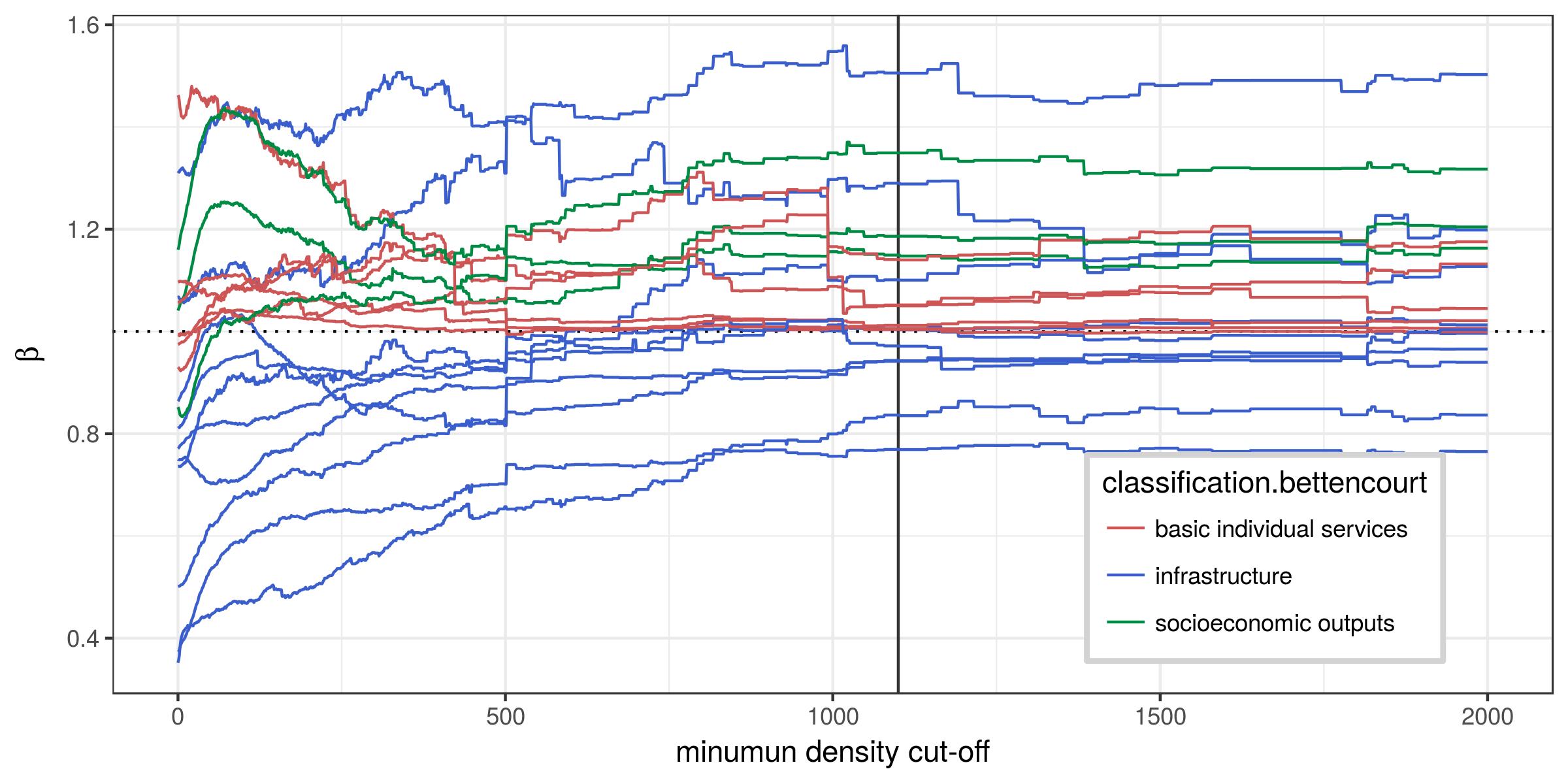}
\end{center}
\end{figure}

\newpage
\paragraph*{S2 Fig}
\label{S2_Fig}

{\bf Exponents values for different urban indicators in the Brazilian urban system}. Each dot represents the scaling exponent related to the best-fit line from the OLS regression of the population against the studied variable; vertical line segments represent 95\% confidence interval (CI) of those regressions; colors are based on the proposed regime; the horizontal black-dotted line indicates linear relationship.
\begin{figure}[ht!]
\begin{center}
\includegraphics[width=1\textwidth]{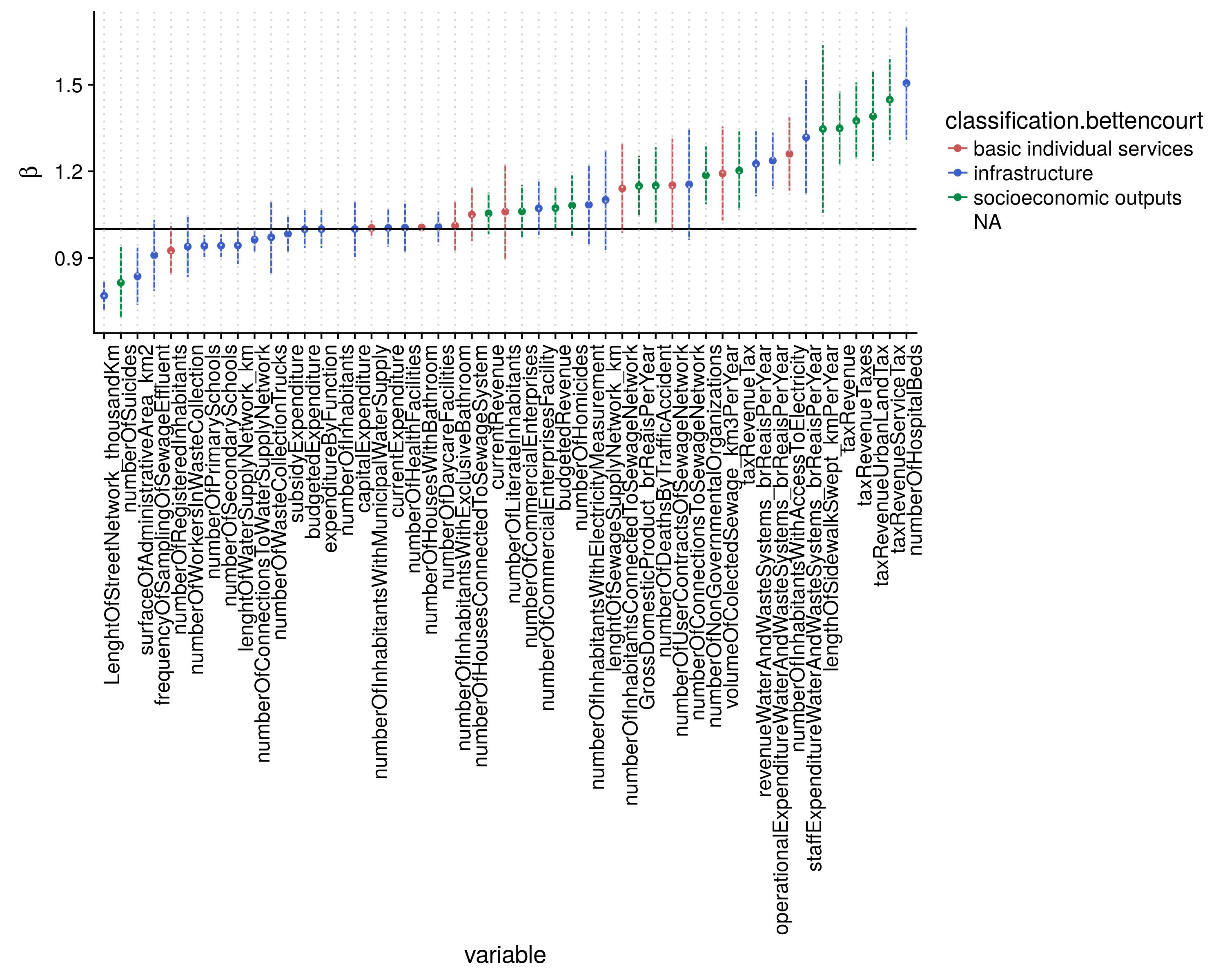}
\end{center}
\end{figure}

\newpage
\paragraph*{S2 Table}
\label{S2_Tab}
{\bf Deviations from the expected scaling regime proposed by Bettencourt \cite{bettencourt2013hypothesis} for all the statistically significant variables.}. Values for $\beta$, and its 95\% confidence interval refers to the final density cut-off value. Statistically insignificant variables are not presented.

\begin{table}[h]
\centering
{\scriptsize 
\begin{tabular}{|l|c|c|}
\hline
{\bf Variable} & {\bf $\beta$}& {\bf Deviation}\\ \thickhline
LenghtOfStreetNetwork &  0.77 [ 0.72 , 0.82 ] &  \\ \hline  
numberOfSuicides &  0.81 [ 0.69 , 0.94 ] & \\ \hline  
surfaceOfAdministrativeArea &  0.84 [ 0.74 , 0.94 ] &  \\ \hline  
frequencyOfSamplingOfSewageEffluent &  0.91 [ 0.79 , 1.03 ] &  \\ \hline  
numberOfRegisteredInhabitants &  0.93 [ 0.84 , 1.01 ] & X \\ \hline  
numberOfWorkersInWasteCollection &  0.94 [ 0.83 , 1.05 ] &  \\ \hline  
numberOfPrimarySchools &  0.94 [ 0.90 , 0.98 ] &  \\ \hline  
numberOfSecondarySchools &  0.94 [ 0.90 , 0.98 ] &  \\ \hline  
lenghtOfWaterSupplyNetwork\_km &  0.94 [ 0.88 , 1.01 ] &  \\ \hline  
numberOfConnectionsToWaterSupplyNetwork &  0.96 [ 0.92 , 1.00 ] &  \\ \hline  
numberOfWasteCollectionTrucks &  0.97 [ 0.84 , 1.10 ] &  \\ \hline  
subsidyExpenditure &  0.98 [ 0.92 , 1.05 ] & X \\ \hline  
budgetedExpenditure &  1.00 [ 0.93 , 1.07 ] & X \\ \hline  
expenditureByFunction &  1.00 [ 0.93 , 1.07 ] & X  \\ \hline
capitalExpenditure &  1.00 [ 0.90 , 1.10 ] & X \\ \hline  
numberOfInhabitantsWithMunicipalWaterSupply &  1.00 [ 0.98 , 1.03 ] &  \\ \hline  
currentExpenditure &  1.00 [ 0.94 , 1.07 ] & X \\ \hline  
numberOfHealthFacilities &  1.01 [ 0.92 , 1.09 ] & X \\ \hline  
numberOfHousesWithBathroom &  1.01 [ 0.99 , 1.02 ] &  \\ \hline  
numberOfDaycareFacilities &  1.01 [ 0.95 , 1.06 ] & X \\ \hline  
numberOfInhabitantsWithExclusiveBathroom &  1.01 [ 0.93 , 1.10 ] &  \\ \hline  
numberOfHousesConnectedToSewageSystem &  1.05 [ 0.96 , 1.14 ] & X \\ \hline  
currentRevenue &  1.05 [ 0.98 , 1.13 ] &  \\ \hline  
numberOfLiterateInhabitants &  1.06 [ 0.90 , 1.22 ] &  \\ \hline  
numberOfCommercialEnterprises &  1.06 [ 0.97 , 1.15 ] &  \\ \hline  
numberOfCommercialEnterprisesFacility &  1.07 [ 0.98 , 1.16 ] & X \\ \hline  
budgetedRevenue &  1.07 [ 1,00 , 1.14 ] &  \\ \hline  
numberOfHomicides &  1.08 [ 0.98 , 1.19 ] &  \\ \hline  
numberOfInhabitantsWithElectricityMeasurement &  1.08 [ 0.94 , 1.22 ] & X \\ \hline  
lenghtOfSewageSupplyNetwork &  1.10 [ 0.93 , 1.27 ] & X \\ \hline  
numberOfInhabitantsServedByWasteCollection &  1.12 [ 1.04 , 1.19 ] & X  \\ \hline  
numberOfInhabitantsConnectedToSewageNetwork &  1.14 [ 0.99 , 1.30 ] & X \\ \hline  
GrossDomesticProduct &  1.15 [ 1.04 , 1.25 ] &  \\ \hline  
numberOfDeathsByTrafficAccident &  1.15 [ 1.02 , 1.28 ] &  \\ \hline  
numberOfUserContractsOfSewageNetwork &  1.15 [ 0.99 , 1.31 ] & X \\ \hline  
numberOfConnectionsToSewageNetwork &  1.15 [ 0.96 , 1.35 ] & X \\ \hline  
numberOfNonGovernmentalOrganizations &  1.19 [ 1.09 , 1.29 ] &  \\ \hline  
volumeOfColectedSewage &  1.19 [ 1.03 , 1.36 ] & X \\ \hline  
taxRevenueTax &  1.20 [ 1.07 , 1.34 ] &  \\ \hline  
revenueWaterAndWasteSystems &  1.23 [ 1.11 , 1.34 ] &  \\ \hline 
operationalExpenditureWaterAndWasteSystems &  1.24 [ 1.14 , 1.33 ] & X \\ \hline  
numberOfInhabitantsWithAccessToElectricity &  1.26 [ 1.13 , 1.39 ] & X \\ \hline  
staffExpenditureWaterAndWasteSystems &  1.32 [ 1.12 , 1.52 ] & X \\ \hline  
lengthOfSidewalkSwept &  1.35 [ 1.06 , 1.64 ] &  \\ \hline  
taxRevenue &  1.35 [ 1.22 , 1.48 ] &  \\ \hline  
taxRevenueTaxes &  1.38 [ 1.24 , 1.51 ] &  \\ \hline  
taxRevenueUrbanLandTax &  1.39 [ 1.24 , 1.55 ] &  \\ \hline  
taxRevenueServiceTax &  1.45 [ 1.31 , 1.59 ] &  \\ \hline  
numberOfHospitalBeds &  1.51 [ 1.31 , 1.70 ] & X \\ \hline 
\end{tabular}
}
\end{table}

\end{document}